# Radar Observations and the Shape of

# Near-Earth Asteroid 2008 EV5

Michael W. Busch
*Department of Earth and Space Sciences, University of California Los Angeles,*
*595 Charles Young Dr. E., Los Angeles CA 90095*
mbusch@ess.ucla.edu

Steven J. Ostro[†], Lance A.M. Benner, Marina Brozovic, Jon D. Giorgini, Joseph S. Jao
*Jet Propulsion Laboratory, California Institute of Technology, Pasadena CA 91109-8099*

Daniel J. Scheeres
*Department of Aerospace Engineering Sciences, University of Colorado at Boulder, 429 UCB, Boulder, CO 80309*

Christopher Magri
*University of Maine at Farmington, 173 High Street, Preble Hall, Farmington, ME 04938*

Michael C. Nolan, Ellen S. Howell, Patrick A. Taylor
*Arecibo Observatory, HC3 Box 53995, Arecibo, PR 00612*

Jean-Luc Margot
*Department of Earth and Space Sciences and Department of Physics and Astronomy, University of California Los Angeles, 595 Charles Young Dr. E., Los Angeles CA 90095*

Walter Brisken
*National Radio Astronomy Observatory, 1003 Lopezville Road, Socorro, NM 87801*

[†] Deceased, 2008 December 15

Direct editorial correspondence and proofs to:

    Michael W. Busch

    595 Charles Young Dr. E., UCLA

    Los Angeles, CA 90095

    mbusch@ess.ucla.edu

*Abstract*

We observed the near-Earth asteroid 2008 EV5 with the Arecibo and Goldstone planetary radars and the Very Long Baseline Array during December 2008. EV5 rotates retrograde and its overall shape is a 400 ± 50 m oblate spheroid. The most prominent surface feature is a ridge parallel to the asteroid's equator that is broken by a concavity 150 m in diameter. Otherwise the asteroid's surface is notably smooth on decameter scales. EV5's radar and optical albedos are consistent with either rocky or stony-iron composition. The equatorial ridge is similar to structure seen on the rubble-pile near-Earth asteroid (66391) 1999 KW4 and is consistent with YORP spin-up reconfiguring the asteroid in the past. We interpret the concavity as an impact crater. Shaking during the impact and later regolith redistribution may have erased smaller features, explaining the general lack of decameter-scale surface structure.

# 1. Introduction

The near-Earth asteroid 2008 EV5 (hereafter EV5) was discovered on 2008 March 4 by the Catalina Sky Survey (Larson et al. 2006). It made an 8.4 lunar distance (0.022 AU, 3.2 million km) Earth approach on 2008 December 23, and was a very strong target for radar observations. This was EV5's closest Earth approach until 2169.

EV5 has low delta-v for rendezvous (the velocity change required to match location and velocity with the object, starting from low Earth orbit), in the lowest ten percent of near-Earth objects (Benner 2010a). In particular, EV5 is a potential target for a human mission to a near-Earth asteroid, with launch windows in 2023 and 2024 (D.F. Landau et al., pers. comm.). EV5's size, shape, surface properties, and rotation state are therefore of great interest.

EV5's absolute magnitude H = 20.0 suggests a diameter within a factor of two of 300 m (assuming an optical albedo between 0.04 and 0.4). Optical and infrared observations suggest that EV5 is a C- (Somers et al. 2008) or X-class object (Reddy 2009) in the SMASSII taxonomy (Bus et al. 2002). Photometry obtained by Galad et al. (2009) and B. Koehn (pers. comm.) shows a rotation period of 3.725 ± 0.001 h and a low lightcurve amplitude, ~0.06 mag, implying that EV5's shape is not elongated.

# 2. Observations and Shape Modeling

## 2.1. Observations

We observed EV5 using the Goldstone 8560-MHz (3.5-cm) radar during 2008 Dec 16-23 and the Arecibo 2380-MHz (12.6-cm) radar during 2008 Dec 23-27 (Table 1). Following a standard protocol (e.g. Ostro et al. 2002, Magri et al. 2007), during each transmit-receive cycle ("run") we transmitted a circularly polarized signal for a time equal to the round-trip light travel time to EV5. For continuous-wave (CW) runs the transmitted signal was a monochromatic sine wave, while for imaging runs a repeating pseudorandom binary code was used to modulate (i.e. flip or not flip) the sinusoid's phase at regular intervals. We then switched to receive mode for an equal time, receiving power in both the circular polarization sense opposite that transmitted (OC) and in the same sense (SC). Single reflections from a smooth surface produce a purely OC echo, whereas multiple reflections and/or diffuse scattering from wavelength-scale structure in the asteroid's near surface yield an echo with nonzero SC/OC circular polarization ratio.

Images were decoded by cross-correlating the received voltage time series with the transmitted code, providing time delay resolution equal to the phase modulation interval. Delay is proportional to distance from Earth (range); delay images spatially resolve the target along the line of sight. The decoded signal in each delay cell was Fourier transformed, providing Doppler frequency resolution; the same was done for each CW spectrum. The center frequency of the radar echo is determined by the target's instantaneous velocity along the line of sight. The target's rotation spreads the echo's frequency about the echo center, providing information about surface elements' line-of-sight velocities that places a joint constraint on the target's shape and spin vector. Images resolve the target in both delay (range) and Doppler; CW spectra provide only Doppler resolution but generally have higher signal-to-noise ratio (SNR) and better calibration than images, thus making them particularly useful for determining the target's radar cross-section and reflectivity.

We first refined our knowledge of EV5's orbit by measuring the echo's offsets from the predicted Doppler shift and time delay, using CW and low-resolution imaging. These ephemeris updates were followed by imaging at the highest delay resolution then available: 0.125 µs (19 m) at Goldstone and 0.05 µs (7.5 m) at Arecibo.

The transmit power of the radars varied widely over the course of our tracks (Table 1). Goldstone used slightly less than half power (205 kW) because only one of the two klystron amplifiers was operational. At Arecibo, the transmit power was ~500 kW during most tracks. On Dec 24, the observatory's generator did not work, forcing the use of commercial power, which reduced the transmitter power to only 67 kW. Even so, EV5's radar echoes were strong enough for high-resolution imaging on all days.

## 2.2. Delay-Doppler Images

The Goldstone images constrain EV5's shape and provide leverage to determine its pole direction. Over the course of the Goldstone tracks, the echo bandwidth changed from 11.5 Hz on 2008 Dec 16 to 14.5 Hz on 2008 Dec 23 (Fig. 1), implying significant change in the object's subradar latitude (the angle between the asteroid-Earth line and the object's equatorial plane). The echo bandwidths of the radar images do not change significantly as EV5 rotates during each day of observation. Thus, the asteroid is not elongated, which is consistent with the lightcurve data. The delay-Doppler images from all four days at Arecibo show prominent structures on EV5's surface (Fig. 2) and the motion of these features agrees with the 3.725-hour rotation period estimated from the lightcurves. The echo extends ~200 m in range, implying a diameter of roughly 400 m if EV5 is approximately spherical. Despite its rounded appearance, EV5 does have significant surface topography: there is a single large concavity (Fig. 3) and an equatorial ridge (Sec. 2.4).

Delay-Doppler radar images are ambiguous: other than at the leading edge of the echo, each point in the northern hemisphere of the object plots in the same position as one in the southern. However, the concavity breaks the echo's leading edge, so that we can measure its overall extent and depth if not its exact shape. The concavity is 150 m across, or ~1/3 of the asteroid's diameter, and occupies roughly 45º of its circumference.  It is roughly 30 m deep and its delay-Doppler structure is suggestive of an impact crater.  Concavities of similar size relative to the object have also been seen in radar images and shape models of asteroids 1998 ML14 (Ostro et al. 2001) and 1998 WT24 (Busch et al. 2008) and in spacecraft images of 2867 Steins (Burchell & Leliwa-Kopystynski 2010), although not on 25143 Itokawa (Fujiwara et al. 2006, Saito et al. 2006).

On scales smaller than the concavity (decameters), EV5's surface appears to be quite smooth.  There are few or no large blocks evident in the 7.5 m-resolution Arecibo images, in contrast to observations of other near-Earth objects such as 2006 VV2 (Benner et al. 2007) and 1998 CS1 (Benner et al. 2009), where small clusters of bright pixels inferred to be large boulders can be tracked as the target rotates.  Despite this lack of decameter blocks, EV5 has a fairly rough surface on the scale of the radar wavelengths (centimeters to decimeters, Sec. 3.1).

*2.3. Shape Modeling*

Using the SHAPE code (Hudson 1993, Magri et al. 2007), we used a constrained least-squares process to estimate EV5's pole direction, shape, and radar scattering properties; accompanied by visual inspection of the fits to verify their quality (Fig. 4).  For the shape modeling, we used the Arecibo CW data and all of the Arecibo images but only every fifth Goldstone image.  The Arecibo images have much higher signal-to-noise ratios and resolution; the Goldstone images are useful for shape modeling primarily in establishing the range extent and bandwidth of the echo and estimating EV5's pole direction rather than the details of the shape.  However, the Goldstone tracks are longer and cover entire rotations of EV5.  Using only a fraction of the Goldstone images significantly reduced the computer time required for the fitting without sacrificing either spatial resolution or rotational coverage of EV5's surface.

Although we recorded images in both circular polarizations, we did not use the SC data in the shape modeling because they are significantly weaker than the OC data.

We began our fits with ellipsoidal shapes and a 20º-resolution grid search of all possible pole directions. We then refined the pole-direction search, with fixed poles spaced at 5º intervals around those directions that matched the observed echo bandwidths, using 10th-order spherical harmonic representations of the shape followed by polyhedra with up to 2000 vertices (for an object of EV5's size, a 2000-vertex model has facets ~20 m across, so further increases in the number of vertices reaches the resolution limit of the data). To prevent SHAPE from fitting non-physical shapes, we included penalty functions to suppress shapes that were excessively oblate or topographically rugged, and to suppress deviations from uniform density and from non-principal axis rotation. Since the penalty functions can also suppress real structure, we ran fits with many different penalty weights and compared the quality of the fits and the corresponding shapes with each other.

Using only the delay-Doppler images, we found two possible pole directions and corresponding shapes – mirror images of each other. The pole directions have J2000 ecliptic longitudes and latitudes of (0º, +84º) and (180º, -84º) ± 10º. The images cover all possible sub-radar longitudes but only equatorial to mid sub-radar latitudes (-10º to +40º for the prograde pole direction, +10º to -40º for the retrograde pole direction), so they do not distinguish between the two pole possibilities. By tracking EV5's radar speckle pattern between the Very Long Baseline Array (VLBA) stations at Pie Town and Los Alamos, Busch et al. (2010) determined that EV5 rotates retrograde. We therefore estimate EV5's pole direction as (180º, -84º) ± 10º, and adopt the corresponding shape model.

We attempted to improve the estimate of EV5's rotation period by including lightcurve data obtained at Modra Observatory between 2008 Dec 24 and 2009 Jan 4 (Galad et al. 2009) in our fits. We tried trial periods spaced at 0.0002 h intervals, and also allowed `SHAPE` to adjust the period simultaneously with EV5's shape and optical scattering properties. However, since EV5's lightcurve amplitude is so low, we cannot separate slight changes in the rotation period from facet-scale adjustments to the shape or albedo changes on the surface. The $3.725 \pm 0.001$ h period from the lightcurves alone remains the best estimate of the rotation period.

Fig. 5 shows our best-fit shape model, and Table 2 lists its physical properties. EV5 is not elongated and an equator-aligned ridge and concavity are required to fit the images (shapes without a ridge give clearly worse fits to the echo behind the leading edge). However, our choice of a preferred shape is somewhat subjective. We adjusted the penalty functions and selected the nominal shape to minimize the overall oblateness (the magnitude of the ridge) while also minimizing the facet-scale structure. This is in keeping with our previous work, where we try to err on the side of avoiding inferring interesting topography that is not actually present on the asteroid at the risk of omitting real structure.

More oblate shapes can give comparably good fits to the data because the images do not cover EV5's north pole. Topography there is primarily constrained by the dynamical and topographic penalty functions: the pole-to-pole extent can be varied by $\pm 50$ m without significantly affecting the fit to the echo's range extent. Reducing the penalty for oblateness depresses the north pole while retaining the equatorial ridge (Sup. Fig. 1) and gives a model that resembles the shape of 2867 Steins (Keller et al. 2010). EV5's equatorial dimensions are constrained to $\pm 50$ m by the bandwidth of the echo. These uncertainties are supported by tests comparing our nominal model and alternate shapes to the Goldstone data that were not used in the fitting: variations smaller than the uncertainties produce comparably good fits.

Decreasing the penalty for facet-scale structure produces a random pattern of 10-m-scale lumps that have no clear signature in the images (SHAPE is fitting noise, Sup. Fig. 1). We therefore chose a relatively high facet-scale penalty weight, sufficient to suppress most decameter structures. Still higher penalty weights begin to sacrifice the quality of the fit to the concavity.

*2.4. The equatorial ridge*

The Arecibo images show relatively strong echo power well behind the leading edge of the echo (Fig. 6), a pattern that indicates that EV5 is not spherical and suggests a ridge aligned with the equator. The leading ridgeline is oriented almost normal to the line of sight and produces a strong radar return. The sides of the ridge maintain a nearly constant incidence angle and hence only a slowly weakening echo before the surface curves away to the poles. A sphere or ellipsoid, or any convex surface, with the same radar scattering law would not produce sufficient echo power behind the leading edge. A convex surface with a more diffuse scattering law could produce echo power at higher incidence angles and hence further back from the leading edge, but then the limbs of the echo, where the incidence angles are also high, would be far brighter than observed. A ridge aligned with the equator produces the correct distribution of echo power as the object rotates.

The delay-Doppler signature of EV5's ridge is much more subtle than that of 1999 KW4 Alpha (the primary of that binary system, Ostro et al. 2006). This is partially because EV5 is a smaller object than KW4 Alpha (400 m vs. 1500 m), but EV5's ridge also has much less relief compared to an ellipsoid approximation of the object's shape (~40 m / 400 m v. ~300 m / 1500 m).  The more subdued ridge on 1994 CC (Brozovic et al., in prep.) is a better comparison to that on EV5.

Because the observations did not include very high sub-radar latitudes, the precise latitude of the ridge is not well constrained.  Depressing one of the poles necessarily moves the ridge off of the equator by displacing the asteroid's center of mass into the other hemisphere, and relatively large changes in the ridge latitude result only in small changes in its distance from Earth.  The ridge can be up to 20º north or south of the equator without significantly decreasing the quality of the fit, and such models were repeatedly found by SHAPE.  The ridge seems likely to lie on the equator, as observed on 1999 KW4 Alpha (Ostro et al. 2006), but we cannot say this definitively.

Our nominal shape is similar to that of 1999 KW4 Alpha and to those of other asteroids observed by radar, such as 1999 RQ36 (Nolan et al. 2007) and 2004 DC (Taylor et al. 2008).  Such a shape suggests a rubble-pile internal structure (Harris et al. 2009).  Oblate shapes with equatorial ridges are characteristically produced on rubble-pile objects that reconfigure due to being spun up (Ostro et al. 2006, Walsh et al. 2008, Harris et al. 2009, Holsapple 2010).

## 3.  Physical Properties of EV5

### 3.1.  Radar scattering properties

EV5's average ratio of SC to OC echo power is 0.40 ± 0.07 at S-band and 0.38 ± 0.02 at X-band (Table 3, Figs. 1 & 7).  This is above average but within the range, 0.28 ± 0.12, of the 17 C-class near-Earth objects that have been observed with radar, and on the low side of the range, 0.67 ± 0.44, of the 5 X-class radar-observed NEOs (Benner et al. 2008).  A higher polarization ratio indicates more multiple scattering of the incident radar beam and a rougher near-surface on decimeter scales. For comparison, SC/OC = 0.28 ± 0.04 for Itokawa; EV5's surface is somewhat rougher.

EV5's radar cross-section is consistent between the Arecibo and Goldstone data and across different rotation phases and days (Table 3). Using the average cross section obtained at Arecibo, we derive a radar albedo of 0.29 ± 0.09 (radar albedo = radar cross-section / geometric cross-section, computed during each epoch of CW observation and averaged). Based on the three different methods described in Magri et al. 2001, we infer a maximum near-surface bulk density of 3.0 ± 1.0 g cm$^{-3}$. EV5's polarization ratio is high enough that the radar albedo – density relationships, which are based solely on single-scattering measurements, are suspect. However, the uncertainty above encompasses the range seen on other objects (e.g. Benner 2010b).

This density range, combined with the optical albedo of 0.12 ± 0.04, is consistent with a range of normal-porosity silicate-carbonaceous and silicate-metal mixtures. We can definitively exclude an E-class enstatite achondrite composition, which would produce a higher optical albedo. Nor is EV5 largely metallic, which would produce a higher radar albedo. Beyond this, we cannot favor one of the reported C and X classifications over the other.

If the bulk density within tens of cm of the surface is representative of EV5's average density, EV5's total mass is (1.0 ± 0.5) x 10$^{11}$ kg, between 1.4 and 4.3 times the mass of Itokawa (Fujiwara et al. 2006).

### 3.2. Pole direction

La Spina et al. 2004 noted an excess of retrograde-rotating objects in the near-Earth population. This excess has been explained as a combination of YORP thermal torques (e.g Bottke et al. 2006) aligning some fraction of asteroid spins to purely retrograde and Yarkovsky-driven migration from the main-belt into resonances with Jupiter and Saturn and then into Earth-crossing orbits favoring retrograde spins (Kryszczynska et al. 2007). EV5's pole direction is within 10º of 180º obliquity, perhaps making it another example of this process.

*3.3. Limits on satellites*

Approximately one-sixth of near-Earth asteroids larger than 200 m in diameter have satellites (Margot et al. 2002), with the fraction rising to about two-thirds for objects with rotation periods less than 2.8 hours (Pravec et al. 2006). For 1999 KW4 (Ostro et al. 2006, Scheeres et al. 2006) and many other binary objects (notably 2004 DC; Taylor et al. 2008) both the primary's equatorial ridge and the satellite are believed to have formed due to YORP spin-up and reconfiguration (Walsh et al. 2008) accompanied by material being shed from the equator. Despite its relatively slow rotation, EV5's shape raises the question of if it has or once had a satellite. We therefore searched the highest-SNR Arecibo images for companions.

Our search covered the asteroid's entire Hill sphere (radius $34 \pm 6$ km based on EV5's heliocentric orbit and our mass estimate) with a resolution of 0.05 µs x 0.0625 Hz (7.5 m x 7.9 mm s$^{-1}$) and integration time ~100 s, but revealed no satellites. This places a limit of ~30 m on any possible companions. A rapidly rotating satellite could be broadened in Doppler frequency so that its echo could not be detected above the noise. However, such a satellite would have to be spinning faster than once every 5 minutes, which seems unlikely due to tidal interactions with the primary. We therefore have confidence in our 30-m limit.

If there are any smaller objects in orbit around EV5, some may be trapped in stationary orbits around equilibrium points above the asteroid's equator. There are two stable equilibria around EV5, along the y-axis of the shape model; ~320 m from the asteroid's center-of-mass or about 90 m above the surface.

## 4. Implications

### 4.1. The equatorial ridge

How did the equatorial ridge form?  It has 40 m of relief (measured vertically to the ridgeline from a sphere fit to the mid-latitude and polar regions) and extends entirely around EV5, with only small irregularities other than the concavity.  If EV5 is a rubble pile, then the ridge presumably formed due to reconfiguration during a time of rapid rotation.  EV5 is not currently spinning quickly enough that we would expect reconfiguration to produce a ridge (for the equator to reach escape velocity, as on 1999 KW4 Alpha, EV5 would have to rotate with a two-hour period).  Figure 8 shows the geopotential estimated using our nominal shape and reveals that the ridge is at higher geopotential than the mid-latitudes.  The geopotential has been expressed as equivalent velocity, which is the velocity of a particle with kinetic energy equal to the gravitational binding energy it would have if it were placed on the surface at that point.  Higher geopotential corresponds to lower equivalent velocity and to a surface less tightly bound to the asteroid.

For bulk densities between 2 and 4 g cm$^{-3}$ the average gravitational slope (the angle between local acceleration and inward normal vectors to each facet) of the shape model is ~13° and the maximum slope is ~35°, along the rim of the concavity (Fig. 9).  The entire surface is below the angle of repose for granular material, with the caveat that our penalties may have suppressed some small regions of high slope.  The ridge's presence therefore suggests that EV5 was spinning more rapidly in the past.

There are several non-exclusive possibilities for previously faster rotation.  If EV5 spun up under the influence of YORP, then perhaps the formation of the ridge was followed by mass being shed into orbit, producing a secondary that tidally evolved outward, slowed EV5's rotation, and has since been lost. A secondary could also have migrated *inward* (Taylor & Margot 2008) and recombined with the object to form a final ridge shape that then spun down due to YORP (Jacobson & Scheeres 2010). Other shape changes – from the concavity's formation or smaller reconfigurations – could have changed the YORP torque on EV5 from net spin-up to net spin-down. For our model's exact shape, YORP should currently be spinning down EV5, with the equator having been at breakup speed ~800 kyr ago (using the method of Scheeres 2007). However, depending on the asteroid's shape below the model's resolution (e.g. Statler 2009), the true YORP torque may be either positive or negative.

### 4.2. The concavity

Is the concavity an impact crater? It has a diameter roughly 1/3 the diameter of EV5, has a depth-to-diameter ratio of about 0.2, is a gravitational low (Fig. 8), and is at least partially surrounded by a rim. This morphology suggests that it is an impact crater. It is possible that EV5's concavity is not a crater, and that that portion of the ridge is instead composed of a few relatively large blocks that happen to have left a gap. However, there are no other depressions of comparable size anywhere on EV5. The relative smoothness at decameter scales also seems to contradict the presence of large blocks. We believe that an impact crater is the most plausible interpretation of the concavity.

The concavity overlies the ridge and therefore must post-date its formation or presumably the process that produced the ridge would have modified the concavity. If the concavity is an impact crater, then models of the seismic activity immediately following an impact predict that many surface features should have been eradicated (Asphaug 2010). However, although EV5's surface is smooth on 10-m scales and there are no other large craters evident in the images, the ridge is still present. A plausible solution is that EV5's internal structure is relatively efficient at dissipating vibrations.  In the framework of Asphaug 2010, it has an attenuation coefficient similar to larger objects such as 433 Eros.  Then the effects of impact-induced shaking should be primarily local and small structures would be erased, while larger structures – namely the ridge – remained.

Craters as large or larger in comparison to the size of the body as EV5's have been seen on many other objects (e.g. Phobos, Veverka et al. 1974; Mathilde, Thomas et al. 1999; Steins, Keller et al. 2010; and possibly 1998 WT24, Busch et al. 2008).  On the main-belt asteroid Mathilde, craters 60% of the object's diameter are one indicator of a high porosity of ~50%.  However, craters one-third of the object diameter do not require high porosity.  For example, Stickney crater is ~1/3 the diameter of Phobos, which has an average porosity of ~30% (Andert et al. 2010).  A similar relative crater diameter and porosity are inferred for 2867 Steins (Burchell & Leliwa-Kopystynski 2010).  Without additional composition information, we cannot estimate EV5's porosity, but an unusually high porosity is not required to explain the concavity, although it is one possible cause of efficient dissipation.

The concavity has the lowest geopotential of any point on EV5's surface (Fig. 8).  Loose regolith may have settled into it and in the mid-latitudes, and away from the ridge; as seen for gravitational lows on Eros (Veverka et al. 2001) and Itokawa (Saito et al. 2006).  Such regolith settling may also have covered up smaller surface features (Richardson 2009).

## 5.  Future Observation Opportunities

With the radar ranging and Doppler astrometry we obtained during our observations and the available ground-based and satellite observations of EV5, its close approaches to Earth can be predicted between 1809 and 2219 (Table 4).  The next opportunities for radar imaging of EV5 occur during its close Earth approaches in 2023 and 2039.  The approach distances during 2023 (0.0422 AU) and 2039 (0.0486 AU) will be more than twice that of the 2008 approach, but EV5 will still be a radar imaging target with estimated signal-to-noise ratios per day of several thousand at Arecibo and several hundred at Goldstone. The sub-radar latitude will extend to +20º as compared to +10º in 2008, improving our knowledge of the north polar region. Other than during these close approaches, EV5 remains at low solar elongation making ground-based optical and near-infrared measurements difficult.

Radar and lightcurve observations during the 2023 and 2039 encounters may determine the current YORP torque on the asteroid and could yield a measurement of the displacement caused by the Yarkovsky effect (Bottke et al. 2006), and thus provide a direct estimate of the asteroid's mass, bulk density, and thermal inertia. Should a mission be sent to EV5 in 2024, observations during the 2023 encounter would provide final trajectory information.

**Acknowledgements**

The Goldstone Solar System Radar, Arecibo Observatory, and NRAO Socorro & Green Bank staff helped to obtain the data presented here. We thank A. Galad, B.W. Koehn, M.D. Hicks, and V. Reddy for providing EV5's rotation period and spectral type from their optical and infrared observations in advance of publication. The Arecibo Observatory is part of the National Astronomy and Ionosphere Center, which is operated by Cornell University under a cooperative agreement with the National Science Foundation. The NRAO is run by Associated Universities, Inc. for the National Science Foundation. Some of this work was performed at the Jet Propulsion Laboratory, California Institute of Technology, under contract with the National Aeronautics and Space Administration (NASA). This paper is based in part on work funded by NASA under the Science Mission Directorate Research and Analysis Programs. M.W. Busch was supported by the Hertz Foundation. Our observations of EV5 are dedicated to the memory of Steven Ostro.

# Table 1: Radar observations

| Time (UTC) | RA,Dec (°) | Dist (AU) | Transmit Power (kW) | Orbit Solution | Resolution (µs x Hz) | # Runs | Sub-radar Position (long start, end; lat) (°) | | |
|---|---|---|---|---|---|---|---|---|---|
| **Goldstone** | | | | | | | | | |
| 2008 Dec 16 13:17–13:26 | | | 205 | | CW | 10 | | | |
| 2008 Dec 16 13:47–13:54 | | | 205 | | 1.0 | 8 | | | |
| **2008 Dec 16 13:58–14:10** | **148 –26** | **0.028** | **205** | **25** | **0.125 x 1.000** | **12** | **190** | **173** | **–42** |
| 2008 Dec 17 09:21–09:30 | | | 205 | | CW | 10 | | | |
| 2008 Dec 17 09:43–09:46 | | | 205 | | 1.0 | 3 | | | |
| **2008 Dec 17 10:00–12:05** | **147 –23** | **0.026** | **205** | **27** | **0.125 x 1.640** | **128** | **59** | **320** | **–39** |
| 2008 Dec 19 08:16–08:23 | | | 205 | | CW | 10 | | | |
| 2008 Dec 19 08:37–08:39 | | | 205 | | 1.0 | 3 | | | |
| **2008 Dec 19 08:56–12:15** | **145 –13** | **0.024** | **205** | **31** | **0.125 x 1.000** | **195** | **332** | **251** | **–31** |
| 2008 Dec 21 07:39–07:47 | | | 205 | | CW | 10 | | | |
| 2008 Dec 21 07:58–08:00 | | | 205 | | 1.0 | 3 | | | |
| **2008 Dec 21 08:16–13:39** | **142 –2** | **0.022** | **205** | **31** | **0.125 x 0.498** | **414** | **1.45 rotations** | | **–20** |
| 2008 Dec 23 09:30–08:37 | | | 205 | | CW | 10 | | | |
| **2008 Dec 23 10:24–12:00** | **139 11** | **0.022** | **205** | **35** | **0.125 x 1.000** | **131** | **163** | **13** | **–9** |
| **Arecibo** | | | | | | | | | |
| 2008 Dec 23 06:51–06:54 | | | 580 | | CW | 5 | | | |
| **2008 Dec 23 06:57–08:19** | **139 10** | **0.022** | **495** | **35** | **0.050 x 0.0625** | **107** | **134** | **1** | **–10** |
| 2008 Dec 24 06:20–06:23 | | | 67 | | CW | 5 | | | |
| **2008 Dec 24 06:25–08:00** | **138 17** | **0.022** | **67** | **35** | **0.050 x 0.0625** | **108** | **31** | **235** | **–4** |
| **2008 Dec 26 05:55–08:17** | **134 29** | **0.022** | **500** | **35** | **0.050 x 0.0625** | **123** | **259** | **40** | **+7** |
| **2008 Dec 27 06:21–07:41** | **133 34** | **0.023** | **470** | **35** | **0.050 x 0.0625** | **84** | **288** | **157** | **+13** |
| **Arecibo + VLBA + Green Bank** | | | | | | | | | |
| **2008 Dec 23 08:24–08:51** | **139 10** | **0.022** | **580** | **35** | **CW — multistatic** | | **353** | **311** | **–10** |

Log of our observations of 2008 EV5. CW data are uncoded transmissions, resolving the target only in Doppler frequency. "Runs" refers to the number of transmit-receive cycles for single-station observations. The transmitter frequency was 8560 MHz at Goldstone and 2380 MHz at Arecibo. Sub-radar longitude and latitude are given for our final shape model, with longitude measured relative to the model's +x axis, increasing towards +y. The Goldstone observations on 2008 Dec 21 covered more than one complete rotation. During the multistatic observations, Arecibo transmitted and the VLBA and Green Bank received. Entries in **bold** were used in our shape modeling.

## Table 2: 2008 EV5 Shape Model

*Pole Direction:*               Ecliptic: **(180°, –84°) ± 10°**
                                RA & Dec: **(7h, –66°) ± 10°**

*Maximum dimensions along principal axes:* **(420 x 410 x 390) ± 50 m**
*DEEVE dimensions:*                         **(415 x 410 x 385) ± 50 m**
*Equivalent diameter:*                      **400 ± 50 m**
*Volume:*                                   **0.035 km$^3$ ± 40%**
*Rotation Period:*                          **3.725 ± 0.001 h**

**OC Radar Albedo:**      0.29 ± 0.09

**Optical Albedo:**       0.12 ± 0.04

Our 2008 EV5 model is polyhedral, with 2000 vertices. The pole direction is given in both ecliptic and right ascension – declination coordinates. "DEEVE" is the dynamically equivalent equal-volume ellipsoid, an ellipsoid with the same volume and moment of inertia ratios as the shape model. Radar albedo is the mean of (radar cross-section)/(model geometric cross-section) computed for each epoch of CW data. The optical albedo was computed from the absolute magnitude H per Pravec & Harris 2007.

## Table 3: 2008 EV5 Radar Cross-Section and Polarization Ratios

| Date | OC Cross-Section (km$^2$) | SC/OC |
|------|---------------------------|-------|
| **Goldstone** | | |
| 2008 Dec 16 | 0.034 ± 0.012 | 0.36 ± 0.02 |
| 2008 Dec 17 | 0.037 ± 0.013 | 0.40 ± 0.02 |
| 2008 Dec 19 | 0.040 ± 0.014 | 0.39 ± 0.02 |
| 2008 Dec 21 | 0.038 ± 0.013 | 0.35 ± 0.02 |
| 2008 Dec 23 | 0.026 ± 0.010 | 0.38 ± 0.02 |
| **Arecibo** | | |
| 2008 Dec 23 | 0.041 ± 0.010 | 0.34 ± 0.09 |
| 2008 Dec 24 | 0.034 ± 0.008 | 0.47 ± 0.12 |

**Average OC Cross-Section (km$^2$):** 0.038 ± 0.007   S–band
                                       0.037 ± 0.006   X–band

**OC Radar Albedo:**      0.29 ± 0.09

**Average Circular Polarization Ratio (SC/OC):** 0.40 ± 0.07  S–band
                                                 0.38 ± 0.02  X–band

Opposite-sense-as-transmitted circular (OC) polarization radar cross-section and polarization ratio (SC/OC) of EV5 measured at Goldstone and at Arecibo. The Goldstone data were processed at 0.2 Hz resolution and the Arecibo data at 0.07 Hz resolution. In both cases, the number of independent Fourier transforms was high enough for the formal errors on the cross-section measurements to be nearly normally distributed. For the cross-section measurements, we have included a 35% uncertainty for Goldstone measurements and 25% for Arecibo due to systematic calibration errors. The calibration uncertainties cancel out for the polarization ratios, so they are dominated by the self-noise of the echo.

## Table 4: 2008 EV5 Close Earth Approaches

| Time (CT) | | | Distance from Earth (AU) | Uncertainty in Distance (AU) | Uncertainty in Time (min) |
|---|---|---|---|---|---|
| 1806 | Dec | 20.87475 | 0.041195 | 0.023776 | 1321.3 |
| 1822 | Dec | 26.44332 | 0.032944 | 0.00829 | 2018.1 |
| 1837 | Dec | 19.92018 | 0.054447 | 0.0024435 | 81.7 |
| 1854 | Jan | 4.99559 | 0.073379 | 0.0039725 | 1777.9 |
| 1868 | Dec | 19.06843 | 0.080763 | 0.000025 | 8.11 |
| 1884 | Dec | 30.80563 | 0.05826 | 0.0041925 | 1466.6 |
| 1899 | Dec | 20.02562 | 0.059495 | 0.001425 | 32.6 |
| 1916 | Jan | 1.37793 | 0.05801 | 0.002234 | 783.14 |
| 1930 | Dec | 20.14227 | 0.080928 | 0.0021715 | 13.55 |
| 1947 | Jan | 6.29407 | 0.076223 | 0.0015915 | 716.95 |
| 1961 | Dec | 20.20018 | 0.052814 | 0.0008975 | 29 |
| 1977 | Dec | 24.78725 | 0.025597 | 0.0002065 | 44.01 |
| 1992 | Dec | 21.73033 | 0.014567 | 0.000003 | 0.83 |
| 2008 | Dec | 23.63674 | 0.0216 | 0 | 0 |
| 2023 | Dec | 20.28654 | 0.042252 | 0.0000065 | 0.33 |
| 2039 | Dec | 28.68228 | 0.048623 | 0.0000255 | 7.3 |
| 2054 | Dec | 18.89239 | 0.09085 | 0.0000235 | 0.03 |
| 2071 | Jan | 6.78568 | 0.08171 | 0.0000245 | 11.15 |
| 2071 | Jun | 13.31193 | 0.09928 | 0.000007 | 19.16 |
| 2085 | Dec | 19.41726 | 0.03758 | 0.000004 | 0.17 |
| 2101 | Dec | 20.06407 | 0.046676 | 0.0000505 | 2.05 |
| 2119 | Jan | 10.61237 | 0.091208 | 0.0001465 | 73.71 |
| 2119 | Jun | 14.7937 | 0.094851 | 0.0000465 | 115.38 |
| 2134 | Dec | 20.11125 | 0.049305 | 0.0002385 | 8.41 |
| 2151 | Dec | 20.88296 | 0.033631 | 0.0009065 | 48.46 |
| 2169 | Dec | 23.15408 | 0.019297 | 0.0047465 | 907.9 |
| 2187 | Jan | 6.66627 | 0.084186 | 0.003716 | 1713.6 |
| 2187 | Jun | 16.29181 | 0.09544 | 0.000555 | 1453.9 |
| 2219 | Dec | 25.1107 | 0.025362 | 0.0055925 | 1065.4 |

Past and future close-Earth approaches by EV5.  Times are given in Coordinate Time decimal dates.  Uncertainties are 3-σ.  The table extends from 1806 to 2219 because outside that interval uncertainties in the times of future approaches exceed 10 days (14400 minutes).

This table was generated using a relativistic n-body extrapolation of JPL orbit solution #91, a weighted least-squares estimate of EV5's heliocentric orbit in the reference frame of the JPL DE405 planetary ephemeris.  Solution 91 is based on 946 optical and 5 radar measurements spanning 2008 March 4 to 2010 April 12.  The post-fit normalized residual RMS (root-reduced chi-square) is 0.41. The trajectory prediction also includes direct solar radiation pressure (Giorgini et al. 2008) and a model of the Yarkovsky acceleration (Vokrouhlicky et al. 2001) acting on EV5, using our model's pole direction, a bulk density of 3 g cm$^{-3}$, an equivalent diameter of 400 m, and a constant thermal conductivity of 0.01 W m$^{-1}$ K$^{-1}$.  By 2219, the encounter distance differs by 0.9-σ (0.0017 AU or 0.66 lunar distances) relative to a prediction that does not include these radiation pressure effects.

**Figure Captions:**

Fig. 1: Continuous-wave (CW) radar echo power spectra of EV5 recorded at Goldstone (a) and

Arecibo (b). Solid lines are the echo power in the opposite-sense-as-transmitted circular polarization (OC); dotted lines are the same-sense-as-transmitted (SC). Echo strength is given in standard deviations of the receiver noise. The vertical scale is the same on each day for the Goldstone spectra, but note the changed scale for the Arecibo data: the echo strength on Dec 24 was much lower than on Dec 23 due to lower transmitter power.

Goldstone spectra are weighted sums of 10 runs each and Arecibo spectra are weighted sums of 5 runs.

Fig. 2: Arecibo delay-Doppler images of EV5 from 2008 Dec 23-27. Resolution is 0.05 μs x

0.0625 Hz (7.5 m x 7.9 mm s$^{-1}$). Within each image, Doppler frequency increases from

left to right and distance from Earth from top to bottom. These images are sums of six

Arecibo runs each and each covers ~7° of rotation phase. The asteroid appears to rotate counter-clockwise. In the collage, time increases from top to bottom and left to right.

Fig. 3: Enlargements of several images from Fig. 2, selected to show the concavity (marked with arrows) and its position on the object. The labels give the observation time and rotation phase (Table 1).

Fig. 4: Delay-Doppler images of EV5 from Goldstone and Arecibo used in our shape modeling,

corresponding synthetic images of the shape model in Fig. 5, and plane-of-sky projections of the model. Time increases from top to bottom and left to right. Arecibo images cover 7° of rotation; Goldstone images cover 5° of rotation and are spaced tens of degrees apart (see text). Delay-Doppler images are oriented as in Fig.1. Plane-of-sky images are oriented with north up and east to the left.

Fig. 5: Principal axis views of our EV5 shape model.  The model is viewed from six orthogonal directions, along its principal axes.  Rotation is around the z-axis, with +z in the direction of the angular momentum vector.  Yellow-shaded regions were seen only at incidence angles >45º or not seen at all.

Fig. 6:  Illustration of the delay-Doppler signature of an equatorial ridge.  The top row gives EV5's radar echo as seen by Arecibo at 2008 Dec 06:27 (left) and synthetic delay-Doppler images for spherical shape models with a best-fit radar scattering law (middle left) and diffuse scattering (middle right) and for our final shape model, which includes an equatorial ridge (right).  Plane-of-sky views of the model shapes are shown at the bottom.  Vertical lines in the top row denote the positions in frequency of the profiles shown in the second row, which plot the echo power at a given frequency as a function of time delay from the top (Earthward) edge of the image.  The spherical shape with best-fit scattering has much less echo power well behind the leading edge than the observed echo does, while the diffuse scattering law has too much power at the limbs of the echo.  In contrast, the ridged model has matches both the observed range extent and the echo limbs.

Fig. 7: Fits to the Arecibo CW data in Fig. 1.b.  Solid lines are the observed OC echo spectra; dotted lines are the fits based on our best-fit shape.

Fig. 8: Geopotential mapped as equivalent velocity over the surface of the EV5 shape model, assuming a bulk density of 3 g cm$^{-3}$ (equivalent velocity = (2 * geopotential)$^{-1/2}$ ). Viewing directions as in Fig. 6.  The equatorial ridge is at higher potential than the mid-latitudes.  Note the gravitational minimum centered on the concavity (outlined by the black dashed line).  The lowest potential point in the model is slightly offset to the north

of the equator, but the true low point may be elsewhere in the concavity (the position of the minimum is sensitive to sub-resolution changes in the shape).

Fig. 9: Facet-scale gravitational slope (angle between the local acceleration and inward normal vectors), mapped across the shape model's surface and viewed from the +x axis. The highest slope point occurs along the eastern edge of the concavity, but the only systematic slope feature is the generally lower slopes in the mid-latitudes relative to the ridgeline and to the poles.

**Supplementary Figure Captions:**

Sup. Fig. 1: Alternate shape models of EV5. *Top*. With a low oblateness penalty and high penalty for facet-scale features, SHAPE extends the ridge and depresses the north pole, moving the ridge from the equator to ~20º north latitude. *Bottom*. With a high oblateness penalty and low facet-scale penalty, SHAPE makes a less prominent ridge and fits individual bright noise pixels with 10-m-scale lumps that are not evident on visual inspection of the images.

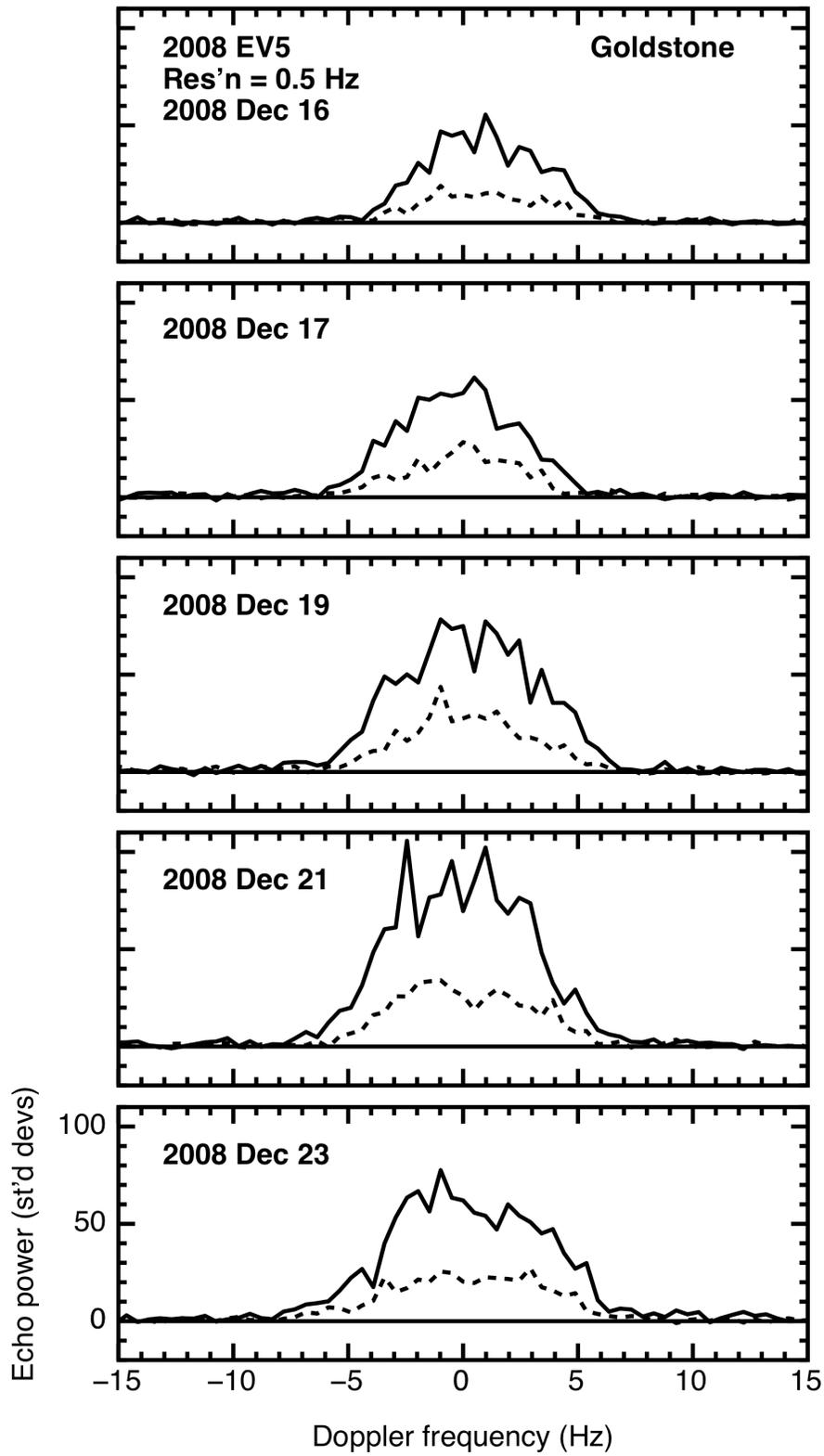

Fig. 1. a.

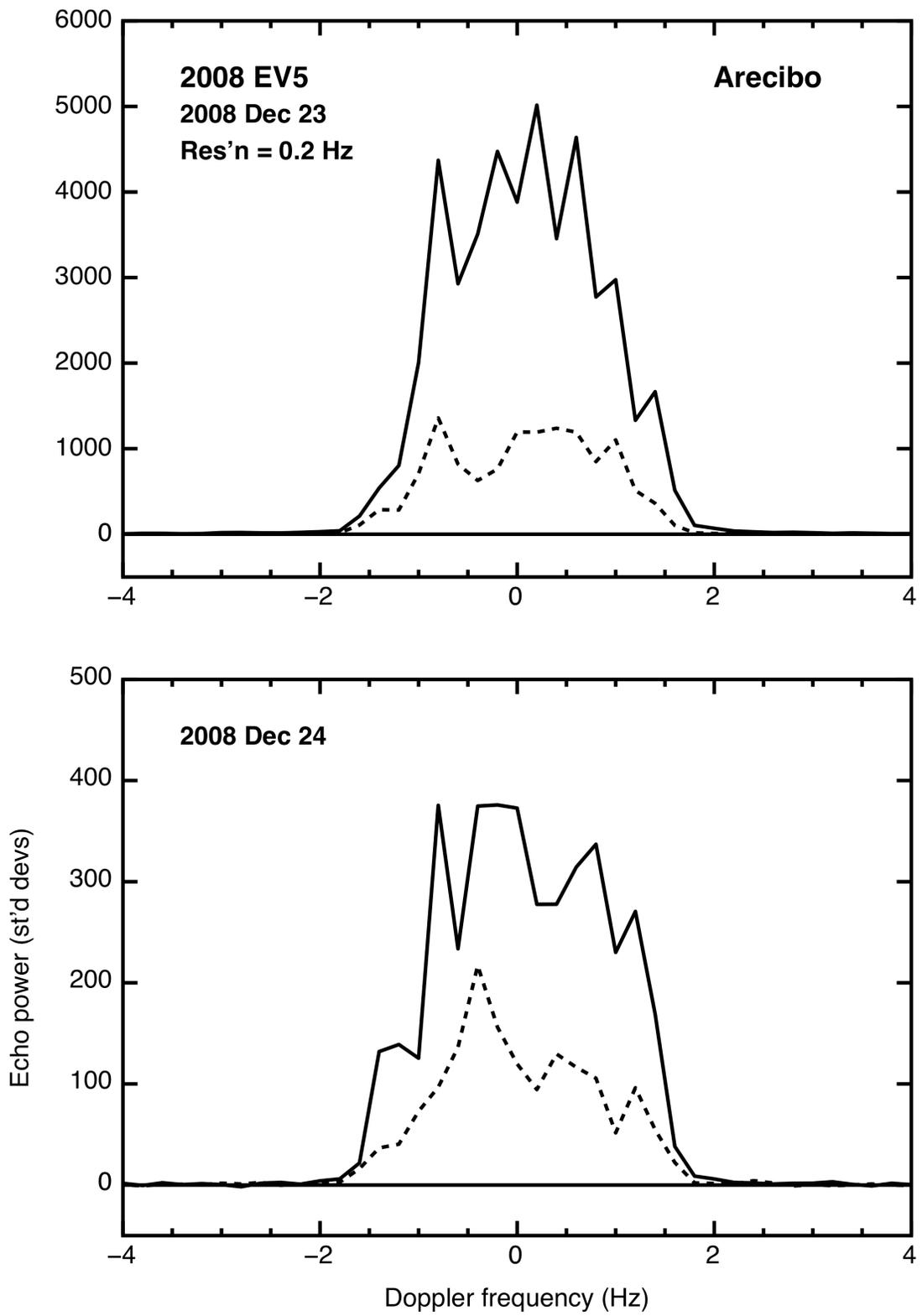

Fig. 1. b.

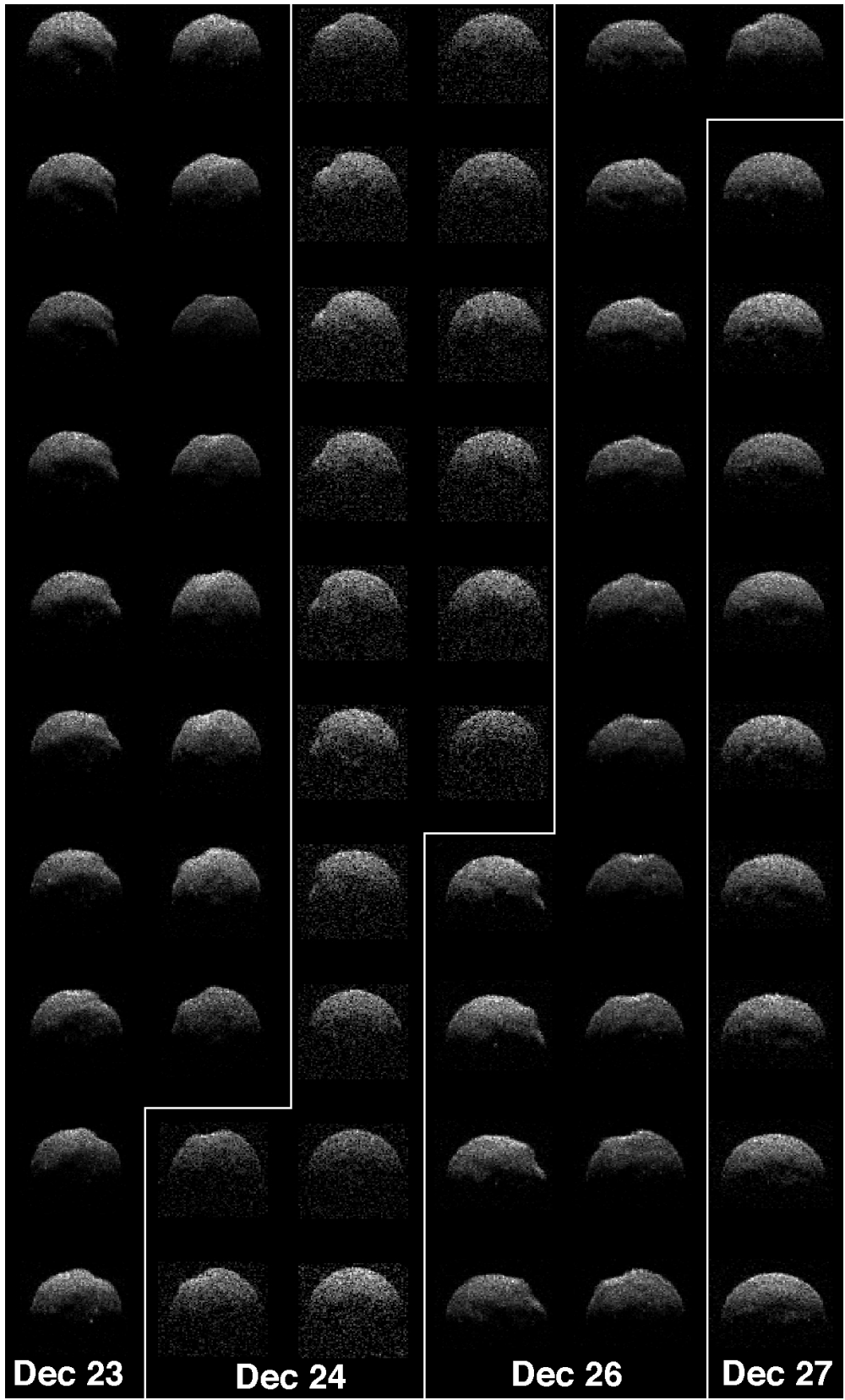

**Dec 23**    **Dec 24**    **Dec 26**    **Dec 27**

Fig. 2

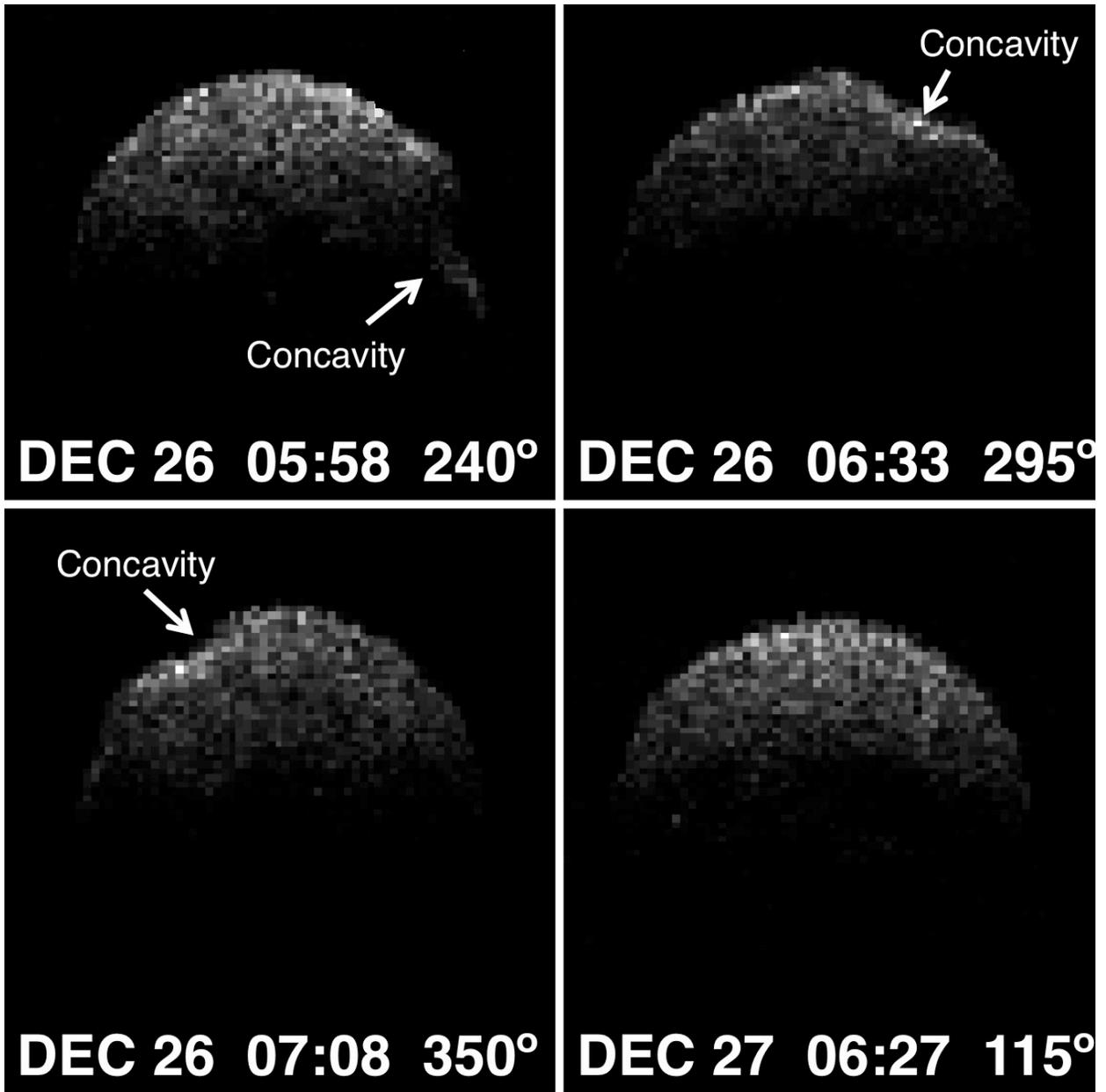

Fig. 3

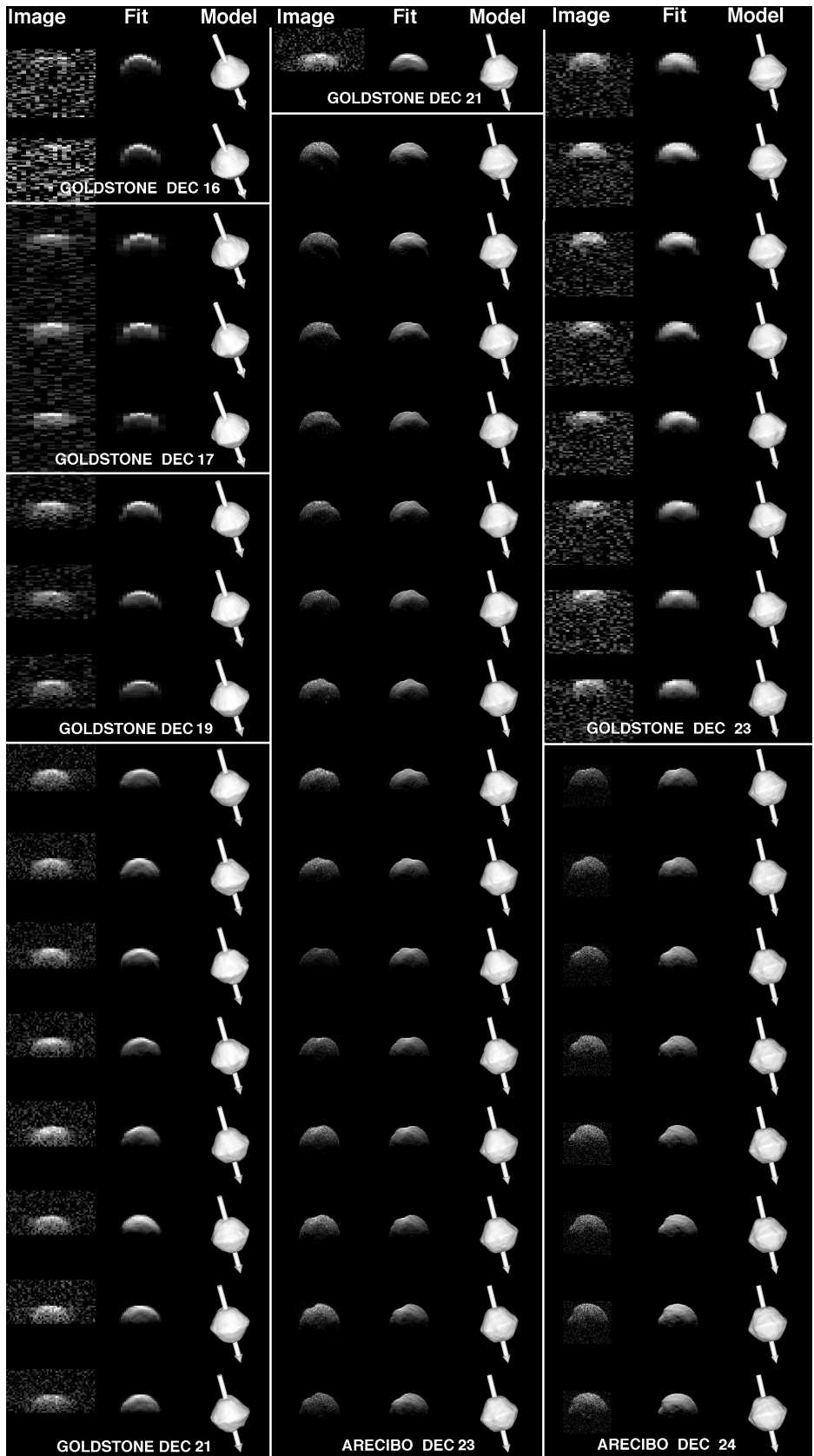

Fig. 4, page 1.

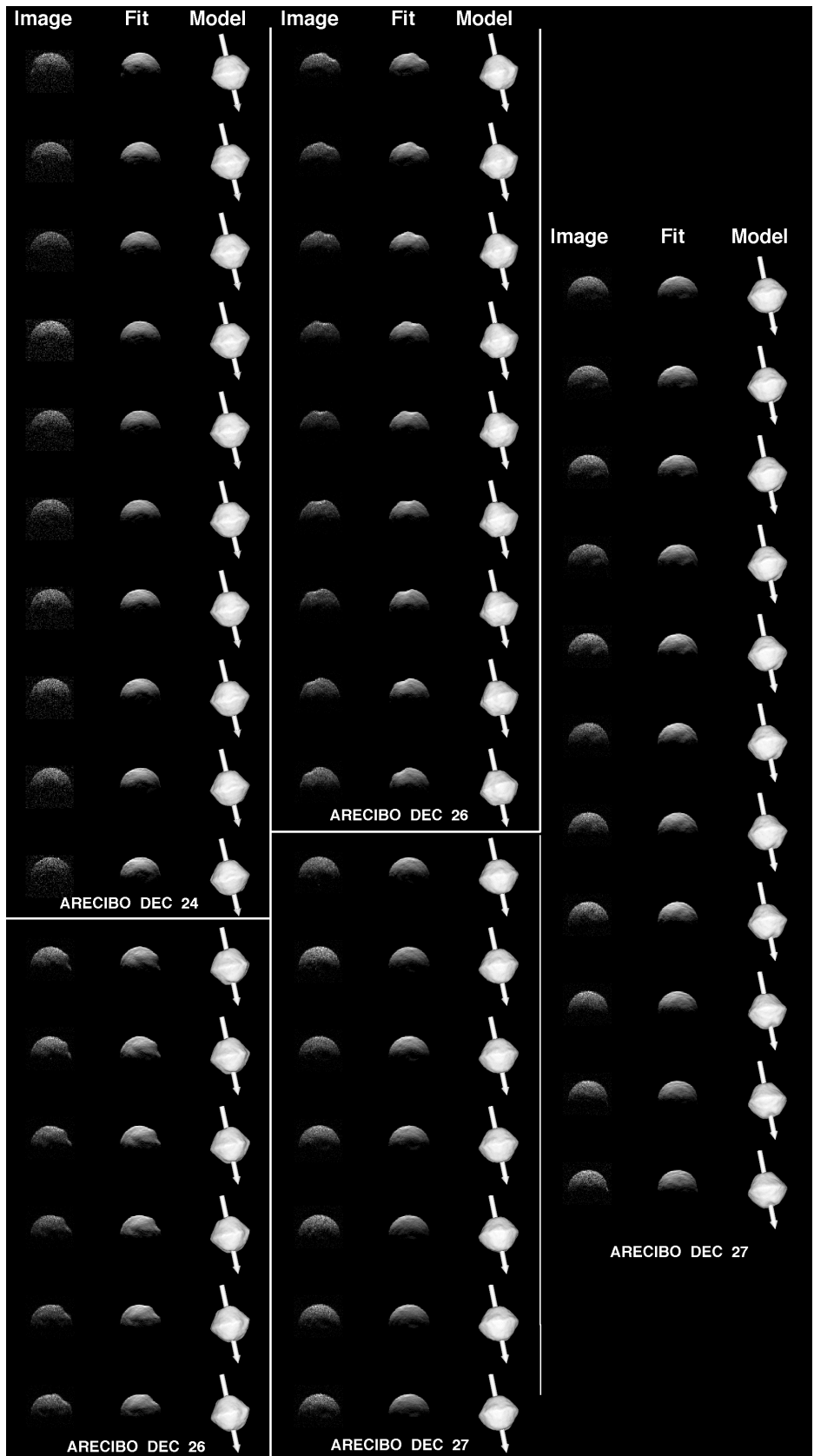

Fig. 4, page 2.

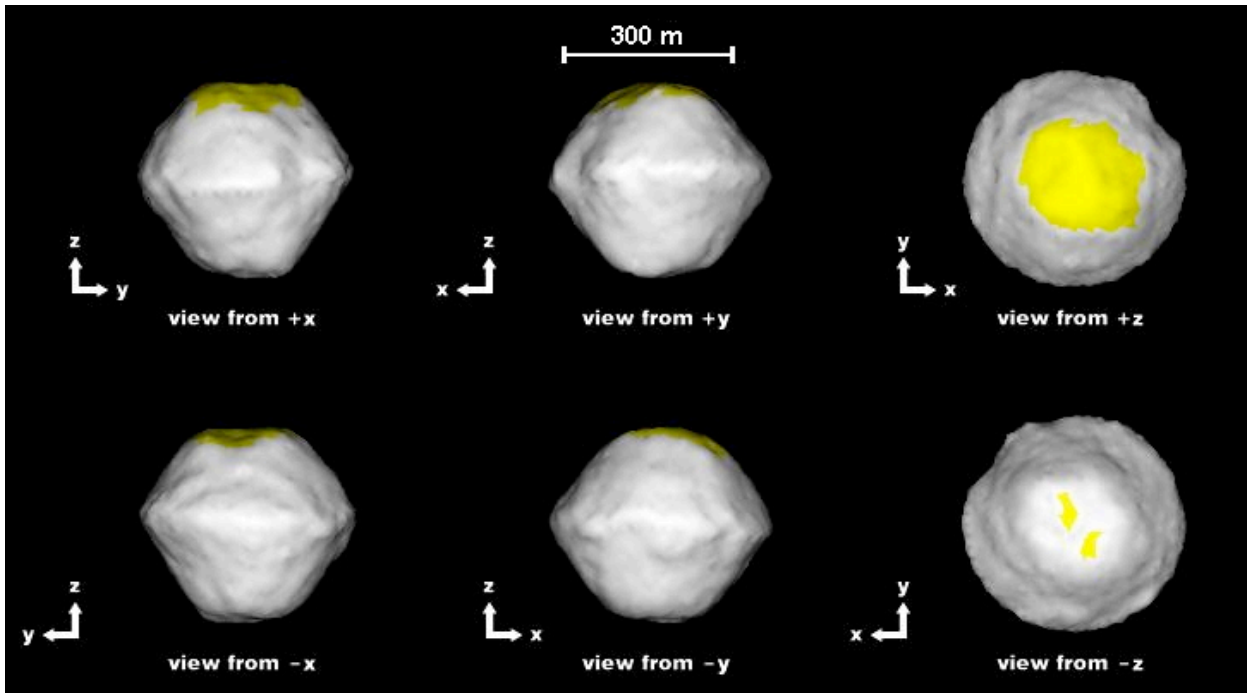

Fig. 5

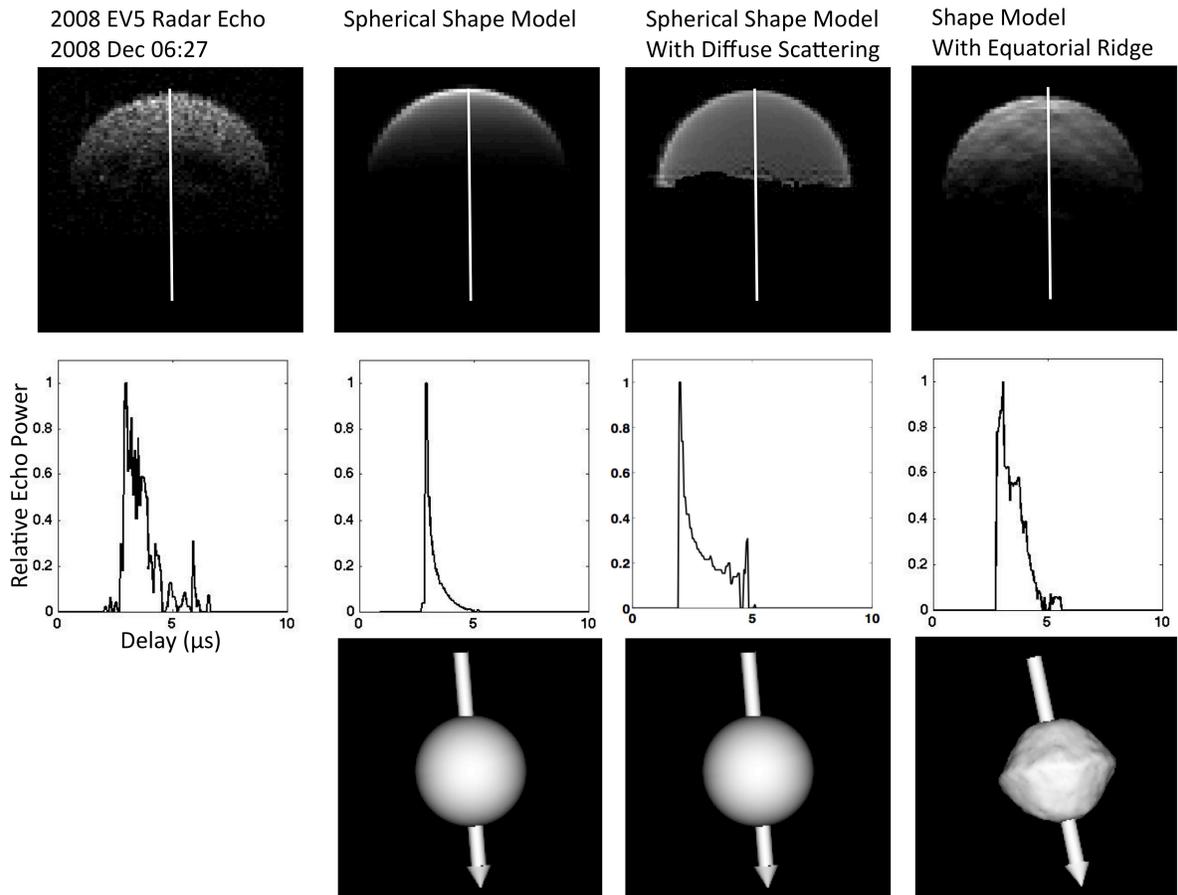

Fig. 6

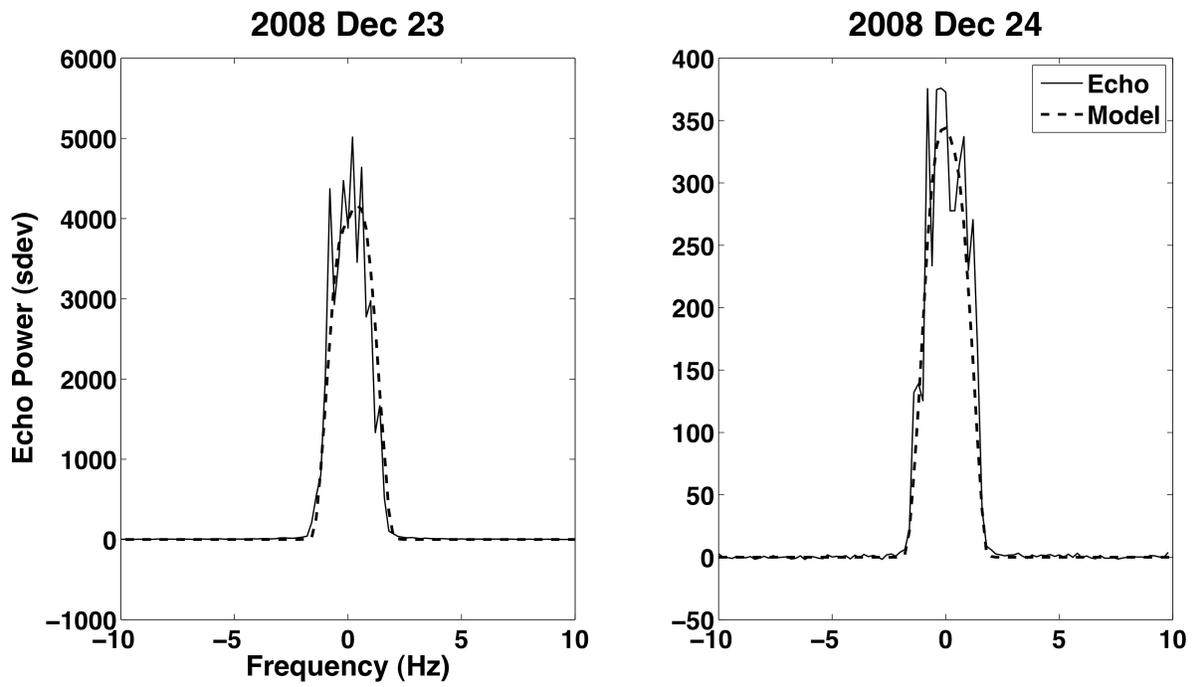

Fig. 7

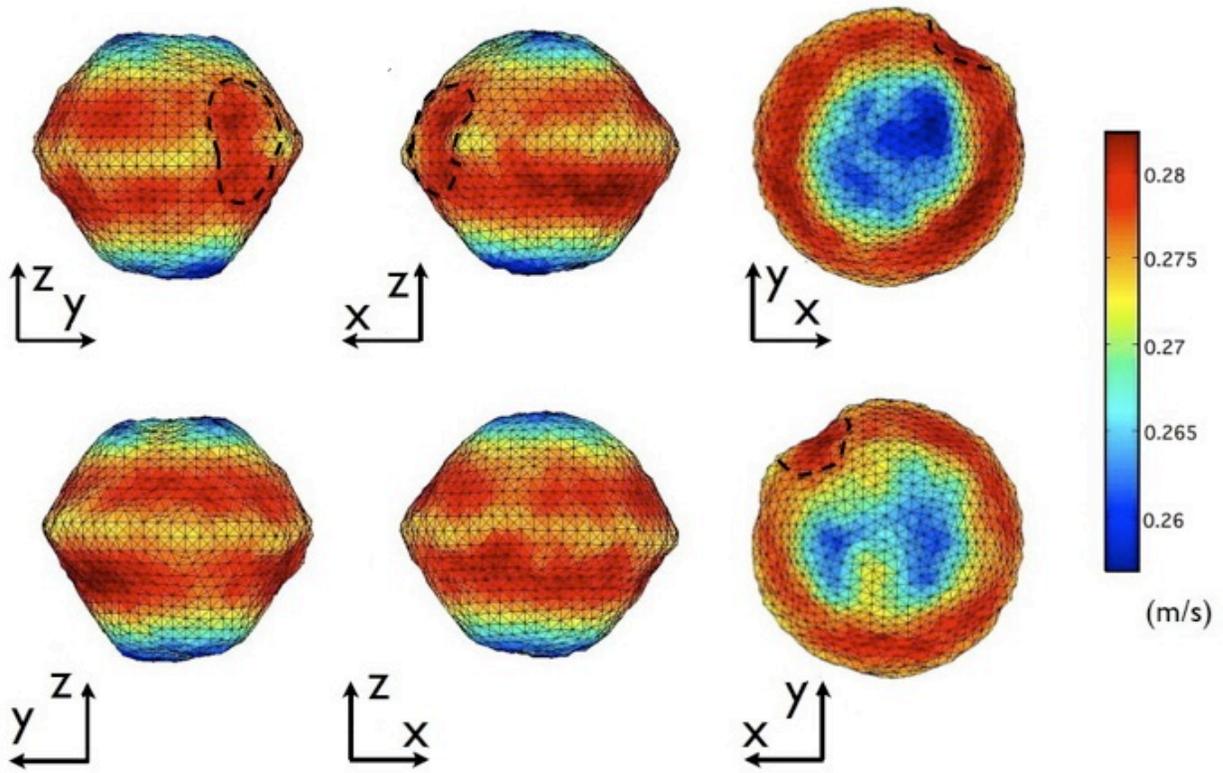

Fig. 8

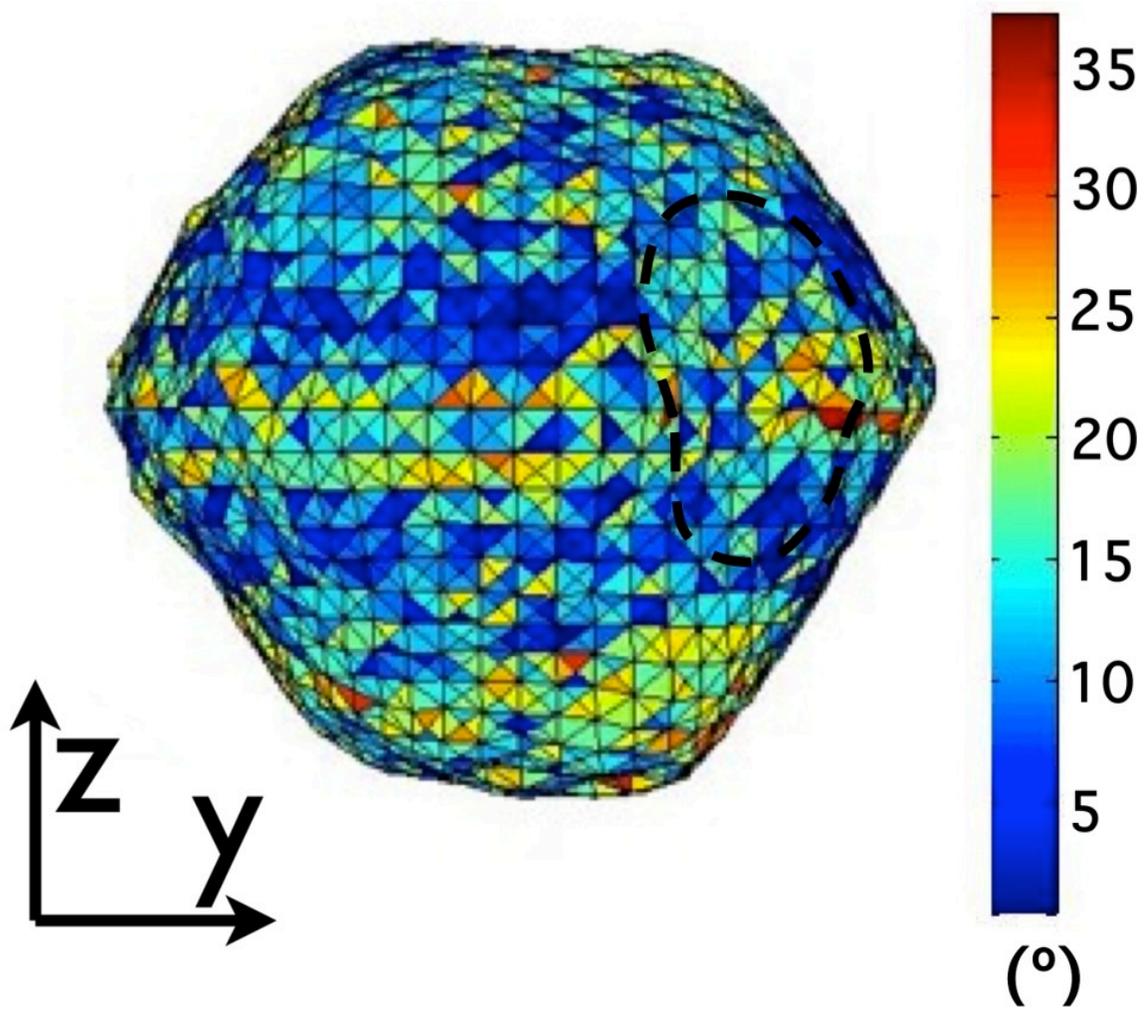

Fig. 9

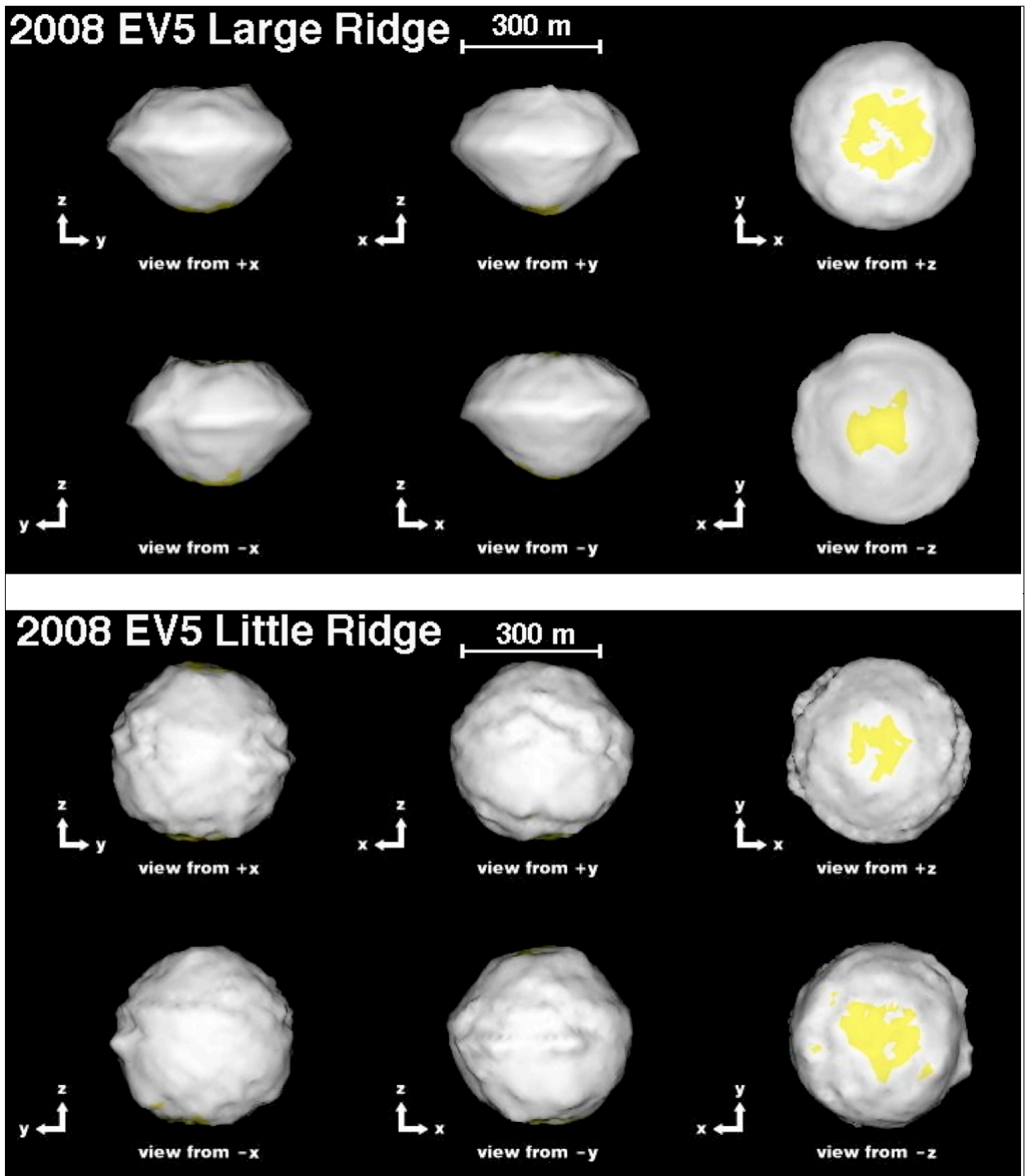

Sup. Fig. 1